\documentclass [12pt,a4paper]{article}

\usepackage{pslatex}
\usepackage{amsfonts,color,morefloats}
\usepackage{amssymb,amsmath,latexsym,amsthm}
\usepackage{bbm}
\usepackage{amsfonts}
\usepackage{mathrsfs}
\usepackage{amsthm}
\usepackage{latexsym}
\usepackage{color}
\usepackage{dcolumn}
\usepackage{extarrows}

\usepackage{amsmath,amssymb,amsbsy,amsfonts,amsthm,latexsym,amsopn,amstext,amsxtra,euscript,amscd,color}
\usepackage{color,xcolor}
\usepackage{amsfonts}
\usepackage{amsfonts,mathrsfs}
\usepackage{psfrag}
\usepackage{subfigure}
\usepackage{url}
\usepackage{stfloats}

\usepackage{latexsym}

\newtheorem{theorem}{Theorem}
\newtheorem{lemma}{Lemma}

\newcommand{\quash}[1]{}

\begin{document}

\def\F{\mathbb{F}}
\def\Z{\mathbb{Z}}

\begin{center}
\textbf{\Large{On $k$-error linear complexity of binary sequences derived from Euler quotients modulo $2p$}}\footnote {The work was partially supported by the Projects of International Cooperation and Exchanges NSFC-RFBR No.~
61911530130, the National Natural Science Foundation of China under grant No.~61772292, by the Provincial Natural Science
Foundation of Fujian under grant No.~2018J01425.
~~ Vladimir~Edemskiy was also partially supported by RFBR and NSFC according to the research project No.~19-55-53003.
~~ Chunxiang  Xu was also partially supported by the National Natural Science
Foundation of China under grant No.~61872060 and the National Key R\&D Program of China under grant No.~2017YFB0802000.\\
}
\end{center}

\begin{center}
\small

Chenhuang Wu$^{a,b}$, Vladimir~Edemskiy$^c$, Chunxiang Xu$^{a}$
\end{center}

\begin{center}
\textit{\footnotesize  a. School of Computer Science and Engineering, University of Electronic Science and Technology of China, Chengdu, Sichuan 611731,  China\\
b. Provincial Key Laboratory of Applied Mathematics, Putian University, Putian, Fujian 351100, China\\
c. Department of Applied Mathematics and Information Science, Novgorod State University, Veliky Novgorod, 173003, Russia\\
\{ptuwch@163.com, vladimir.edemsky@novsu.ru, chxxu@uestc.edu.cn\}
}
\end{center}


\noindent\textbf{Abstract:} We consider the $k$-error linear complexity of binary sequences derived from Eluer quotients modulo $2p$ ($p>3$ is an odd prime), recently introduced by J. Zhang and C. Zhao. We adopt certain decimal sequences to determine the values of $k$-error linear complexity for all $k>0$. Our results indicate that such sequences have ¡®good¡¯ stability from the viewpoint of cryptography.

\noindent\textbf{keywords:} cryptography, pseudorandom binary sequences, Euler quotients, $k$-error linear complexity

\section{Introduction} \label{Introduction}
In recent years, Fermat quotients \cite{EM1997,OS} and Euler quotients \cite{AD1997, DKC12} have been widely used in the design of pseudorandom sequences. Research indicates that such sequences have ¡®good¡¯ cryptographic and pseudorandom properties \cite{Chen, CD, COW, GW, WJL2015}. In the research of using Euler quotients to construct pseudorandom sequences, these works mainly concentrate on the modulus which are of odd prime power \cite{CDM2015, CEKW2018,  DCH12, DKC12}. Recently, J. Zhang and C. Zhao designed binary sequences with period $2p^2$ by using the Euler quotients modulo $2p$, they proved that such sequences had very high linear complexity \cite{ZZ2018}. A related construction had been considered from Euler quotients modulo $2p$ in an earlier work \cite{WCD2012}. Very recently, R. Mohammed et al. extended the work of \cite{ZZ2018} to $r$-ary sequences and discussed their linear complexity \cite{MDL2019}.

For an integer $m>1$, the Euler quotients $q_m(u)$ is defined as
\begin{equation}\label{Eq}
q_m(u)=\frac{u^{\phi(m)}-1}{m}\pmod m,\,\ 0 \leq q_m(u)<m,
\end{equation}
where integers $u\geq 0$ with $\gcd(m,u)=1$ and $\phi(\ )$ is the Euler totient function\cite{AD1997}. We also set $q_m(u)=0$ for $\gcd(m,u)\neq 1$.

When $m$ is an odd prime, Eq.(\ref{Eq}) is just the Fermat quotients \cite{EM1997,OS}. In the literature, Carmichael quotients are also studied by using the Carmichael function, the reader is referred to \cite{M2015}.

Then, when $m=2p$ for an odd prime $p$, J. Zhang and C. Zhao \cite{ZZ2018} considered the following binary sequence $(e_u)$:
\begin{equation}\label{sequence}
e_u=\left\{
\begin{array}{ll}
0, & \mathrm{if}\ \ 0 \leq \frac{q_{2p}(u)}{2p} < \frac{1}{2},\\
1, & \mathrm{if}\ \ \frac{1}{2}\leq \frac{q_{2p}(u)}{2p} < 1,
\end{array}
\right.  ~~ u \geq 0,
\end{equation}
which is $2p^2$-periodic. They proved that the linear complexity satisfied
\begin{equation}\label{LC}
LC((e_u))=\left\{
\begin{array}{ll}
2(p^2-p), & \mathrm{if}\,\ p\equiv 1 \pmod 4, \\
2(p^2-1), & \mathrm{if}\,\ p\equiv 3 \pmod 4,
\end{array}
\right.
\end{equation}
if $2^{p-1}\not \equiv 1 \pmod{p^2}$.

It is well known that any periodic sequence can be reproduced by linear feedback shift register (LFSR) \cite{CDR}. The linear complexity is an important cryptographic measure of sequences. The Berlekamp-Massey algorithm \cite{B1968,M1969} can effectively recover the entire sequence from a subsequence of length
twice of its linear complexity by LFSR. Therefore, the linear complexity of a sequence must be larger than half of its period. In the following, we review the linear complexity of periodic sequences.

Let $\F$ be a field.  For a $T$-periodic
sequence $(s_n)$ over $\F$, recall that the
\emph{linear complexity} over $\F$, denoted by  $LC^{\F}((s_n))$, is the smallest positive integer $L$ such that
$$
s_{n+L} = c_{L-1}s_{n+L-1} +\ldots +c_1s_{n+1}+ c_0s_n\quad
\mathrm{for}\,\ n \geq 0,
$$
which is satisfied by $(s_n)$ and where $c_0\neq 0, c_1, \ldots,
c_{L-1}\in \mathbb{F}$.
Let
$$
S(X)=s_0+s_1X+s_2X^2+\ldots+s_{T-1}X^{T-1}\in \mathbb{F}[X],
$$
which is called the \emph{generating polynomial} of $(s_n)$. Then the linear
complexity over $\F$ of $(s_n)$ can be computed as
\begin{equation}\label{licom}
  LC^{\F}((s_n)) =T-\deg\left(\gcd(X^T-1,
  ~S(X))\right),
\end{equation}
see, e.g., \cite{CDR, MN2002} for details.

We know that, for a sequence to be cryptographically strong, its linear complexity should be high. However, this complexity also should not significantly be reduced by changing a few terms which leads to the notion of the $k$-error linear complexity \cite{SM}( also see \cite{DXS} for the related sphere complexity that was defined even earlier). For integers $k\ge 0$, the \emph{$k$-error linear complexity} over $\F$ of $(s_n)$, denoted by $LC^{\F}_k((s_n))$, is the lowest linear complexity (over $\F$) that can be
obtained by changing at most $k$ terms of the sequence per period.  Clearly, $LC^{\F}_0((s_n))=LC^{\F}((s_n))$, and
$$
T\ge LC^{\F}_0((s_n))\ge LC^{\F}_1((s_n))\ge \ldots \ge LC^{\F}_w((s_n))=0
$$
where $w$ equals the number of nonzero terms of $(s_n)$ per period.

The organization of this work is as follows. Section \ref{Introduction} reviews some definitions needed in the work. Section \ref{Polynomials} gives two results on the polynomials of degree $<2p^2$ over $\F_2$, a binary field. Section \ref{Lemmas} presents some lemmas that will be used in the proof of our main results. Section \ref{Results} discusses the $k$-error linear complexity of the recently proposed binary sequences derived from Euler quotients. Section \ref{Examples} lists two examples. Finally, Section \ref{Conclusion} concludes the work.

\section{Some results of polynomials of degree $<2p^2$}\label{Polynomials}

Because the period of binary sequences discussed in this work is $2p^2$, from Eq.(\ref{licom}) we only need to consider the polynomials with degree $<2p^2$ over $\F_2$. Let
\begin{eqnarray*}
\Phi_1(X)&=&X-1,\\
\Phi_2(X)&=&1+X+\ldots+X^{p-1},\\
\Phi_3(X)&=&1+X^p+X^{2p}+\ldots+X^{(p-1)p}.
\end{eqnarray*}

Therefore, we can get that $$X^{2p^2}-1=\big(X^{p^2}-1\big)^2=\Phi_1(X)^2\Phi_2(X)^2\Phi_3(X)^2.$$ If 2 is a primitive root modulo $p^2$, then $\Phi_1(X), \Phi_2(X), \Phi_3(X)$ are irreducible polynomials over $\F_2$ \cite{LN-FF}.

Let
$$
S(X)=s_0+s_1X+s_2X^2+\ldots+s_{2p^2-1}X^{2p^2-1}\in \F_2[X].
$$

We define vectors $$U_i=(s_i,s_{i+2p},\ldots,s_{i+2(p-1)p} ), 0\leq i <2p,$$
and
$$V_j=(s_j+s_{j+p^2},s_{j+p}+s_{j+p^2+p},\ldots,s_{j+(p-1)p}+s_{j+p^2+(p-1)p}), 0\leq j <p.$$
The vectors $U_i$ and $V_j$ also can be considered as the decimal sequences of the first period of $(s_n)$.
Let $wt(\ )$ denotes the Hamming weight of a vector, i.e., the number of nonzero elements in the vector.

Then, we have the following two Lemmas.

\begin{lemma}\label{P32}
$\Phi_3(X)^2\mid\mathcal{S}(X)$ if and only if $wt(U_i)\in \{0,p\}$, for $0 \leq i <2p$.
\end{lemma}
Proof. If $\Phi_3(X)^2|\mathcal{S}(X)$, then there exits a polynomial $h(X)=h_0+h_1X+h_2X^2+\ldots+h_{2p-1}X^{2p-1}\in \F_2[X]$ such that $\mathcal{S}(X)=\Phi_3(X)^2h(X)$.
We note that $\Phi_3(X)^2=1+X^{2p}+X^{4p}+\ldots+X^{2(p-1)p}$. Therefore, if $h_i=0$, then $wt(U_i)=0$. Otherwise, for $h_i=1$, then $wt(U_i)=p$.

If $wt(U_i)\in \{0,p\}$, for $0 \leq i <2p$, it is clear that $\Phi_3(X)^2\mid\mathcal{S}(X)$. We finish the proof. \qed

\begin{lemma}\label{P3}
$\Phi_3(X)\mid\mathcal{S}(X)$ if and only if $wt(V_j)\in \{0,p\}$, for $0 \leq j <p$.
\end{lemma}
Proof. We note that $\Phi_3(X)\mid X^{p^2}-1$. Therefore, $\Phi_3(X)\mid\mathcal{S}(X)$ if and only if $\Phi_3(X)\mid(\mathcal{S}(X)\pmod {X^{p^2}-1}).$ Since $$\mathcal{S}(X)\equiv\sum\limits_{j=0}^{p-1}X^{j}\left(\sum\limits_{l=0}^{p-1}(s_{j+lp}+s_{j+p^2+lp})X^{lp}\right)\pmod {X^{p^2}-1},$$
then we have $\Phi_3(X)\mid\mathcal{S}(X)$ if and only if $\sum\limits_{l=0}^{p-1}(s_{j+lp}+s_{j+p^2+lp})X^{lp}=0$ or $\sum\limits_{l=0}^{p-1}(s_{j+lp}+s_{j+p^2+lp})X^{lp}=\Phi_3(X),$ which derives $wt(V_j)\in \{0,p\}$, for $0 \leq j <p$. We finish the proof. \qed

\section{Structure of the proposed binary sequence}\label{Lemmas}

In this section, we consider the binary sequence $(e_u)$ defined in Eq.(\ref{sequence}). For convenience,  we introduce some vectors whose elements are in $(e_u)$. Denote\\
$A_i=\left(e_i,e_{i+2p},\ldots,e_{i+2(p-1)p} \right), \textmd{where }0\leq i <2p$.\\  $B_j=\left(e_j,e_{j+p},\ldots,e_{j+(p-1)p},e_{j+p^2},e_{j+p+p^2},\ldots,e_{j+(p-1)p+p^2} \right)$,\\
$C_j=\left(e_j,e_{j+p},\ldots,e_{j+(p-1)p}\right)$, $D_j=\left(e_{j+p^2},e_{j+p+p^2},\ldots,e_{j+(p-1)p+p^2} \right)$, i.e., $B_j=\left(C_j,D_j\right)$, where $0\leq j <p$. \\
Let $E_j=C_j+D_j$, where `+' denotes the addition of two vectors over the finite field $\F_2$.

Firstly, we give some lemmas which are necessary for proving our main results.
\begin{lemma}\label{EQ-2p}
For integers $u,v$ with $\gcd(uv,2p)=1$, we have \\
(1). $q_{2p}(u)$ is even,\\
(2). $q_{2p}(u+2tp)=q_{2p}(u)+t(p-1)u^{-1} \pmod {2p}$,\\
(3). $q_{2p}(uv)=q_{2p}(u)+q_{2p}(v) \pmod {2p}$.
\end{lemma}
Proof. See \cite{ZZ2018} for (1) and see \cite{M2015} for (2) and (3). \qed
\textsl{}\begin{lemma}\label{even}
\{$q_{2p}(u+2tp)\mid 0\leq t <p, \gcd(u,2p)=1\}=\{2\ell\mid 0\leq \ell<p$\}.
\end{lemma}

Proof. By Lemma \ref{EQ-2p}(2), for $0\leq t_1,t_2 <p$ and $\gcd(u,2p)=1$, if $q_{2p}(u+2t_1p)=q_{2p}(u+2t_2p)$, then $q_{2p}(u)+t_1(p-1)u^{-1} \equiv q_{2p}(u)+t_2(p-1)u^{-1} \pmod {2p}$.  That is $t_1(p-1)u^{-1} \equiv t_2(p-1)u^{-1} \pmod {2p}$. We get $t_1 = t_2$. Thus, \{$q_{2p}(u+2tp)\mid 0\leq t <p, \gcd(u,2p)=1\}$ exactly contains $p$ elements.

Then by Lemma \ref{EQ-2p}(1) we complete the proof. \qed

\begin{lemma}\label{wt-A}
For $0 \leq i <2p$, we have
\begin{eqnarray*}
wt(A_i)
=\left\{
\begin{array}{ll}
\frac{p-1}{2}, & \mathrm{if}\,\ \gcd(i,2p)=1, \\
0, & \mathrm{otherwise}.
\end{array}
\right.
\end{eqnarray*}

\end{lemma}
Proof. The Lemma follows from Eq. (\ref{sequence}) and Lemma \ref{even}.  \qed

\begin{lemma}\label{wt-BE}
For $0 \leq j <p$, we have\\
(1). $wt(B_j)=wt(C_j)+wt(D_j)=wt(E_j)$,\\
(2). $wt(B_0)=0$ and $wt(B_j)=\frac{p-1}{2}$, for $1 \leq j<p.$
\end{lemma}
Proof. (1).  It is easy to see that $wt(B_j)=wt(C_j)+wt(D_j)$ because of $B_j=\left(C_j,D_j\right)$ for $0 \leq j <p$.

Note that $j+tp$ and $j+tp+p^2$ are of different parity for $0 \leq t<p$. Then, from Eq.(\ref{Eq}) and Eq.(\ref{sequence}), for odd $j: 0\leq j<p$ we have
\begin{equation*}\label{e+e}
e_{j+tp}+e_{j+tp+p^2}=\left\{
\begin{array}{ll}
e_{j+tp}, & \textmd{if}\,\ t \,\ \textmd{is even},\\
e_{j+tp+p^2}, & \textmd{if}\,\ t \,\ \textmd{is odd}.
\end{array}
\right.
\end{equation*}
We have a similar result for even $j: 0\leq j<p$. Thus, $wt(E_j)=wt(B_j)$ since $E_j=C_j+D_j.$

(2). Firstly, since $q_{2p}(u)=0$ for $\gcd(2p,u)\neq 1$, we can easily get $wt(B_0)=0$.

Secondly, for odd $j: 0\leq j<p$, we have $q(j+(2t+1)p)=0$ for $0 \leq t <p$ and
$\{q_{2p}(j+2tp)\mid 0\leq t <p\}=\{2\ell\mid 0\leq \ell<p\}$ by Lemma \ref{even}. Thus, by Eq. (\ref{sequence}), we get $wt(B_j)=\frac{p-1}{2}$.

Thirdly, for even $j: 0\leq j<p$, we have $wt(B_j)=\frac{p-1}{2}$ similarly.  \qed

Clearly, by Lemma \ref{wt-BE}, there are $\frac{(p-1)^2}{2}$ many $1$s in each period of $(e_u)$.

\section{Main results: $k$-error linear complexity}\label{Results}
In this section, we consider the $k$-error linear complexity of $(e_u)$ defined in Eq.(\ref{sequence}). The generating polynomials of $(e_u)$ is denoted as
$$
\mathcal{G}(X)=e_0+e_1X+e_2X^2+\ldots+e_{2p^2-1}X^{2p^2-1}\in \mathbb{F}_2[X].
$$
Let $\mathcal{E}_k(X)\in \mathbb{F}_2[X]$ be a polynomial of degree $<2p^2$ with exactly $k$ many monomials, and
\begin{equation}\label{Hk}
\mathcal{G}_k(X)=\mathcal{G}(X)+\mathcal{E}_k(X)\in \mathbb{F}_2[X].
\end{equation}
Indeed, $\mathcal{G}_k(X)$ is the generating polynomial of the sequence obtained from $(e_u)$ by changing exactly $k$ terms of $(e_u)$ in the first period and continued periodically. Therefore, to discuss the $k$-error linear complexity of $(e_u)$, we only need to compute $\gcd(X^{2p^2}-1,  ~\mathcal{G}_k(X))$ by Eq.(\ref{licom}).

Let $A_i(X)=\sum\limits_{t=0}^{p-1}e_{i+2tp}X^{i+2tp}$ and  $E_j(X)=\sum\limits_{t=0}^{p-1}(e_{j+tp}+e_{j+p^2+tp})X^{j+tp}$, for $0\leq i<2p, 0\leq j<p$.
Then, we have
\begin{equation}\label{G(x)A}
\mathcal{G}(X)=\sum\limits_{i=0}^{2p-1}A_i(X)=\sum\limits_{i=0}^{2p-1}\sum\limits_{t=0}^{p-1}\left(e_{i+2tp}X^{i+2tp}\right),
\end{equation}
and
\begin{equation}\label{G(x)E}
\mathcal{G}(X)\equiv\sum\limits_{j=1}^{p-1}E_j(X)\pmod{X^{p^2}-1}.
\end{equation}

\begin{theorem}\label{klc-p1}
Let $(e_u)$ be the binary sequence of period $2p^{2}$ defined by Euler quotients modulo $2p$ in Eq.(\ref{sequence}), where $p>3$ is an odd prime. If $p\equiv 1 \pmod 4$ and $2$ is a primitive root modulo $p^2$, then
the $k$-error linear complexity of $(e_u)$  satisfies
\[
 LC^{\F_2}_k((e_u))=\left\{
\begin{array}{cl}
2(p^2-p), & \mathrm{if}\,\ 0\le k<(p-1)^2/2, \\
0, & \mathrm{if}\,\ k\geq (p-1)^2/2.
\end{array}
\right.\\
\]
\end{theorem}
Proof. From Eq.(\ref{LC}), we know that $LC_0^{\F_2}((e_u))=2(p^2-p)$ when $p\equiv 1 \pmod 4$ and $2$ is a primitive root modulo $p^2$.

Now, we want to find a polynomial $\mathcal{E}_k(X)$ with the least number $k$ such that $\Phi_3(X)\mid\mathcal{G}_k(X)$. If so, by Eq.(\ref{Hk}), Eq.(\ref{G(x)E}) and Lemma \ref{P3}, $\mathcal{E}_k(X)$ should be of the following form
$$\mathcal{E}_k(X)=\sum\limits_{j=1}^{p-1}F_j(X)\pmod{X^{p^2}-1},$$
where $F_j(X)\in \{E_j(X), X^j\Phi_3(X)-E_j(X)\}$ for $1\leq j<p.$ Clearly, each $F_j(X)$ has $\frac{p-1}{2}$ or $\frac{p+1}{2}$ many monomials. So
$\mathcal{E}_k(X)$ at least contains $\frac{(p-1)^2}{2}$ many monomials. Then we choose $\mathcal{E}_k(X)=\mathcal{G}(X)$ which leads to $\mathcal{G}_k(X)=0$, where the least $k=\frac{(p-1)^2}{2}$. It also means that changing $\frac{(p-1)^2}{2}$ many terms in one period of $(e_u)$ will get a zero sequence, and changing any $k<\frac{(p-1)^2}{2}$ many terms will lead to $\Phi_3(X)\nmid \mathcal{G}_k(X)$.

At the same time, if $\Phi_3(X)^2\mid \mathcal{G}_k(X)$, by Lemmas \ref{P32} and \ref{wt-A}  we need
$$\mathcal{E}_k(X)=\sum\limits_{\stackrel{i=0}{\gcd(i,2p)=1}}^{2p-1}A_i(X),$$
in which case the least $k=\frac{(p-1)^2}{2}$. Therefore, we get for $k<\frac{(p-1)^2}{2}$
$$\min \deg\left(\gcd(X^{2p^2}-1, ~\mathcal{G}_k(X))\right)=2(p^2-p).$$
We complete the proof. \qed

\begin{theorem}\label{klc-p3}
Let $(e_u)$ be the binary sequence of period $2p^{2}$ defined by Euler quotients modulo $2p$ in Eq.(\ref{sequence}), where $p>3$ is an odd prime. If $p\equiv 3 \pmod 4$ and $2$ is a primitive root modulo $p^2$, then the $k$-error linear complexity of $(e_u)$  satisfies

\[
 LC^{\F_2}_k((e_u))=\left\{
\begin{array}{cl}
2(p^2-1), & \mathrm{if}\,\ k=0, \\
2(p^2-p+1), & \mathrm{if}\,\ 1 \leq k<p-1, \\
2(p^2-p), & \mathrm{if}\,\ p-1 \leq k<(p-1)^2/2, \\
0, & \mathrm{if}\,\ k\geq (p-1)^2/2.
\end{array}
\right.\\
\]

\end{theorem}
Proof. From Eq.(\ref{LC}), for $p\equiv 3 \pmod 4$ and $2$ is a primitive root modulo $p^2$, we know that $LC_0^{\F_2}((e_u))=2(p^2-1)$.

Firstly, as the proof in Theorem \ref{klc-p1}, if we want to make $\Phi_3(X)\mid \mathcal{G}_k(X)$, it must change at least $k=\frac{(p-1)^2}{2}$ many terms in each period of $(e_u)$.

Secondly, we consider $\mathcal{G}(X)$ modulo $X^{2p}-1$. We find that by Lemma \ref{wt-A}
$$\mathcal{G}(X)\equiv\sum\limits_{\stackrel{i=0}{\gcd(i,2p)=1}}^{2p-1}X^i \pmod {X^{2p}-1}.$$

If we choose a polynomial $\mathcal{E}_k(X)$ such that $\mathcal{E}_k(X)=X^p \pmod {X^{2p}-1}$, in which case the least $k$ is 1, then we get
$$\mathcal{G}(X)+\mathcal{E}_k(X)=\mathcal{G}(X)+X^p=X\Phi_2(X)^2 \pmod {X^{2p}-1},$$
and $\mathcal{G}(1)+\mathcal{E}_k(1)=\mathcal{G}(1)+1= p \neq 0.$ Thus, $LC_1^{\F_2}((e_u))=2(p^2-p+1)$.

If we choose a polynomial $\mathcal{E}_k(X)$ such that
$$\mathcal{E}_k(X)=\sum\limits_{\stackrel{i=0}{\gcd(i,2p)=1}}^{2p-1}X^i \pmod {X^{2p}-1},$$
in which case the least $k$ is $p-1$, then we get
$$\mathcal{G}(X)+\mathcal{E}_k(X)=0 \pmod {X^{2p}-1},$$
i.e., $(X^{2p}-1)\mid \mathcal{G}_k(X).$
At the same time, we consider $\mathcal{G}(X)$ modulo $X^{p}-1$. We find that by Lemma \ref{wt-BE}(2)
$$\mathcal{G}(X)\equiv\sum\limits_{i=1}^{p-1}X^i \pmod {X^{p}-1}$$
and we need to choose $\mathcal{E}_k(X)\equiv\sum\limits_{i=1}^{p-1}X^i \pmod {X^{p}-1}$ to make $(X^{p}-1)\mid \mathcal{G}_k(X).$ In fact, in this case
$k\geq p-1.$ According above, we have $LC_k^{\F_2}((e_u))=2(p^2-p)$ for $p-1 \leq k <\frac{(p-1)^2}{2}$.
Indeed from above, any $\mathcal{E}_k(X)$ with $1 < k <p-1$ will not make $\Phi_2(X)\mid \mathcal{G}_k(X)$, so $LC_k^{\F_2}((e_u))=2(p^2-p+1)$ for $1 \leq k <p-1$. We finish the proof. \qed

\section{Examples}\label{Examples}
We also run a program to confirm our theorems. The experimental data are listed below and the results are consistent with Theorem \ref{klc-p1} and Theorem \ref{klc-p3}.

\textbf{Example 1.} Let $p=5$ (i.e., $p\equiv 1\pmod 4$, 2 is a primitive root modulo 25). In this case, from Eq.(\ref{sequence}) we can get the sequence $(e_u)$:
\begin{equation*}\label{e+e}
e_{u}=\left\{
\begin{array}{ll}
1, & \textmd{if}\,\ u\pmod{50}=3, 9, 13, 21, 29, 37, 41, 47,\\
0, & \textmd{otherwise}.
\end{array}
\right.
\end{equation*}

That is, the first period of the sequence $(e_u)$ is\\
\hspace{-0.55cm}00010, 00001, 00010, 00000, 01000, 00001, 00000, 00100, 01000, 00100.

The $k$-error linear complexity of the sequence $(e_u)$ is
\[
 LC^{\F_2}_k((e_u))=\left\{
\begin{array}{cl}
40, & \mathrm{if}\,\ 0\le k<8, \\
0, & \mathrm{if}\,\ k\geq 8.
\end{array}
\right.\\
\]

\textbf{Example 2.} Let $p=11$ (i.e., $p\equiv 3\pmod 4$, 2 is a primitive root modulo 121). In this case, from Eq.(\ref{sequence}) we can get the sequence $(e_u)$:
\begin{equation*}\label{e+e+e}
e_{u}=\left\{
\begin{array}{ll}
1, & \textmd{if}\,\ u\pmod{242} \in U,\\
0, & \textmd{otherwise}.
\end{array}
\right.
\end{equation*}
where $U=\{5, 7, 13, 15, 17, 21, 25, 39, 45, 47, 51, 53, 59, 61, 63, 65, 75, 79, 83, 85, 89,\\ 101, 107, 109, 117, 125, 133, 135, 141, 153, 157, 159, 163, 167, 177, 179, 181, 183, 189,\\ 191, 195, 197, 203, 217, 221, 225, 227, 229, 235, 237\}.$

That is, the first period of the sequence $(e_u)$ is\\
\hspace{-0.55cm}00000101000, 00101010001, 00010000000, 00000010000, 01010001010, 00001010101, 00000000010, 00100010100, 01000000000, 00100000101, 00000001000, 00001000000, 01010000010, 00000000001, 00010100010, 00100000000, 01010101000, 00101000101, 00000100000, 00000000100, 01000101010, 00001010000.

The $k$-error linear complexity of the sequence $(e_u)$ is
\[
 LC^{\F_2}_k((e_u))=\left\{
\begin{array}{cl}
240, & \mathrm{if}\,\ k=0, \\
222, & \mathrm{if}\,\ 1 \leq k<10, \\
220, & \mathrm{if}\,\ 10 \leq k<50, \\
0, & \mathrm{if}\,\ k\geq 50.
\end{array}
\right.\\
\]

\section{Conclusion}\label{Conclusion}
   In this work, we considered the $k$-error linear complexity of recently proposed sequences with period $2p^2$ derived from Euler quotients modulo $2p$. The results indicated that such sequences had good stability. We also illustrated our results by two examples.

\section*{Acknowledgements}

Parts of this work were written during a very pleasant visit of Chenhuang Wu to the Novgorod State
University of Russia in 2019. He wishes to thank the host for the hospitality.



\end{document}